# Redundancy Analysis of the Railway Network of Hungary

B. G. Tóth
*National University of Public Service, Budapest, Hungary*

ABSTRACT: Available alternative routes on which traffic can be rerouted in the case of disruptions are vital for transportation networks. Line sections with less traffic under normal operational conditions but with increased importance in the case of disruptions are identified in the railway network of Hungary by using a weighted directed graph. To describe the goodness of the individual alternative routes the so-called redundancy index is used. The results show that the structure of the network is good, but the lines with the highest redundancy (lines No. 80, 2, 4 and 77 according to the numbering of the national railway operator, MÁV) are mostly single tracked and in many cases the line speed is low. The building of additional tracks and electrifying these lines while still maintaining the existing diesel locomotives for the case of disruptions of the electric support are the keys to make the performance of the rather dense railway network of Hungary sustainable.

## 1 INTRODUCTION

Describing a network in its fullness with a single measure is basically impossible. The choice of the measure always has to depend on what kind of property one wants to expose in the light of a specific application. If the global performance is in question, the indices of (Kansky 1989) is a good place to start as they are topology measures of the graph representing the network. In spite of their simplicity, they are effectively used even in network design (Derrible & Kennedy 2011).

To describe individual network elements based on their various roles in the network, the centrality measures are a simple but comprehensive set of indices (Lin & Ban 2013) which can also be weighted with passenger exposure (Cats et al. 2016). Recently, multi-criteria decision analysis has been also used for network element ranking (Almoghatawi et al. 2017).

But is a specific network robust against the disruptions of its most vital elements (Derrible & Kennedy 2010, Laszka et al. 2012)? Can it perform under the same supply and demand pattern even if its most important elements are disrupted (Snelder et al. 2012; Disser & Matuschke 2017)? What happens if the capacity of a link is reduced (Cats & Jenelius 2016) or even if rerouting is necessary (Oliveira at al. 2016)? Identifying these network elements is fundamental for critical infrastructure protection.

But not only the highly threatened has to be defended. The Southern Railway Bridge at Budapest is closed for pedestrians and cyclists and is guarded by armed personnel, because it is the most critical element of the railway network of Hungary. Its smallest disruption affects the whole network as the total international freight transport passing through Hungary crosses the Danube via this bridge. However, the tracks and the other small bridges on the same line section are not defended in any way though any accident or other disruption can happen there and can block the line section.

One thus must be aware of the alternative routes that can be used in the case of the disruptions of the most important network elements and the line sections along these routes also have to be continuously maintained to be real alternatives for the rerouted paths. The railway network of Hungary is the fourth densest in the world (UNECE 2019), which hints that with relatively little cost good possibilities for alternative routes can be established.

In this paper, the lines that would help in keeping the system performance at sufficient level even in the case of the disruption of its most vital elements are identified by using a mathematical model. To identify these line sections, the so-called redundancy measure will be used which quantifies how important a line section in the case of the disruption of other line sections.

## 2 THE GRAPH MODEL OF THE RAILWAY NETWORK OF HUNGARY

To model the railway network of Hungary, a weighted graph was used. The nodes of the graph represented the stations and the edges represented the line sections between the stations. The weights of the edges were either the length of the corresponding line sections or the time needed to pass through them, the latter calculated as the ratio of the length and the line speed of the line sections. The length and permissed speed values are accessible publicly at the web page of the Hungarian Rail Capacity Allocation Office (Hungarian: Vasúti Pályakapacitás-elosztó Kft.) (VPE 2019). Where the line speed was different for trains with locomotives and for EMUs, the lower value was used. This means that the travel times of the model are lower bounds for the real travel times.

The model did not include the stops with no switches, i.e., where reversing of the trains is not possible. Furthermore, nodes with exactly two neighbors, the so-called "joint nodes" (Jenelius et al. 2006) were transformed out: the joint node and the two edges connecting to it were deleted and the two neighboring stations were connected with one edge weighted with the sum of the two deleted edges. The only exceptions were the border crossings and the stations preceding them not to eliminate the border effect fully.

For the reversings, an additional 15-minute increase in the travel time was assigned. When calculating trip lengths, no increase was assigned to reversings. Passing through a station did not increase the travel time or the trip length. To model the increase in the travel times, each station was represented by four nodes instead of one (Fig. 1). At both "sides" of the station an arrival and a departure node was defined and the arrival and departure node on the same sides were connected with a 15-minute edge and the arrival and departure node on the opposite sides with a 0-minute edge. For neighboring stations, the departure node of a station was connected to the arrival node of the neighboring station.

However, this is still not enough: as to make the 15-minute edges not surpassable, the edges had to be directed in a proper way for the paths to arrive at a station in an arrival node and to depart from it from a departure node (Fig. 1).

Wyes were represented similar to stations but because reversing on a wye is not possible without entering the respective station, the edges connecting the arrival and the departure node on the same sides were omitted.

In the case of termini, only one arrival and one departure node with a 15 minute edge between them is enough for this representation.

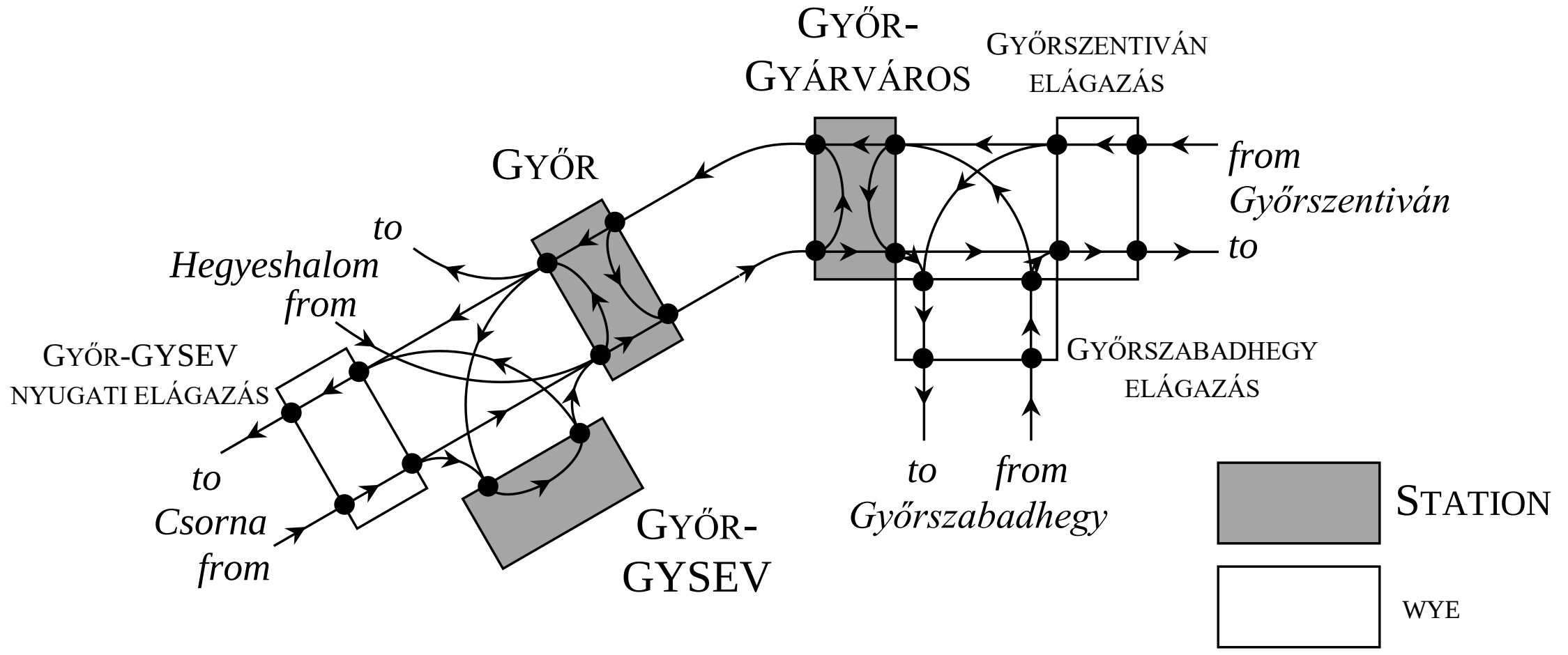

Figure 1. An illustration of the graph model on station Győr and its neighboring stations.

The graph on which the calculations were performed contained a total of 1136 nodes representing 291 stations and 26 wyes and 1808 edges, representing 366 line sections (and the stations and the wyes). To every edge, two weights could be assigned: one length and one travel time value. For one calculation, naturally, only one kind of weight was used. Let us denote the graph weighted with lengths by $G_\ell^0 = (V, E, W_\ell)$ and the graph weighted with travel times by $G_t^0 = (V, E, W_t)$, where $V$ is the set of nodes, $E$ is the set of edges and $W_\ell$ and $W_t$ are the sets of length and travel time weights.

# 3 CALCULATION METHODS

The calculations and their visualization was carried out in the *R* programming language and environment (R Core Team 2012) by using the *igraph* package (Csardi & Nepusz 2006).

## 3.1 *Calculating the shortest path*

Since the distance and travel time values are both positive real numbers and the graph is relatively small, the easiest way to determine the shortest paths between the pairs of stations is Dijkstra's algorithm. The algorithm is implemented in the *igraph* package in the distances() function.

By calculating the shortest path in distance and in time for all $\langle a,b\rangle$ pairs of stations ($a \neq b$) one gets two sets of 42.195 values, which number will be denoted by $N^0$. Let us denote the duration of the fastest path between stations $a$ and $b$ in the $G_t^0$ graph by $t_{ab}^0$ and the length of the shortest path between them in the $G_\ell^0$ graph by $\ell_{ab}^0$.

## 3.2 *Disruption of line sections*

The term "disruption" will be used for the complete closure of a line section. Let us denote the set of edges of the disrupted line section(s) by $e$ ($e \subseteq E$). Let the graph not containing the edges of $e$ be denoted by $G^e = (V, E \setminus e, W^e)$ the weights being either $W_t^e \subseteq W_t$ or $W_\ell^e \subseteq W_\ell$ for time and distance weights of the $G_t^e$ and $G_\ell^e$ graphs, respectively.

# 4 REDUNDANCY

If a shortest path passes through line section $e$ in the undisrupted network, on the disruption of line section $f$ there are two scenarios which are irrelevant for our calculations. First, if the shortest path passes through line section $e$ in the disrupted network, too, then the disruption has no effect on $e$ as it is still useable. Second, if the disrupted line section, $f$, makes at least one pair of stations unreachable for each other, then there is no possible alternative path (see Fig. 2). Therefore, these two scenarios are left out of the calculations.

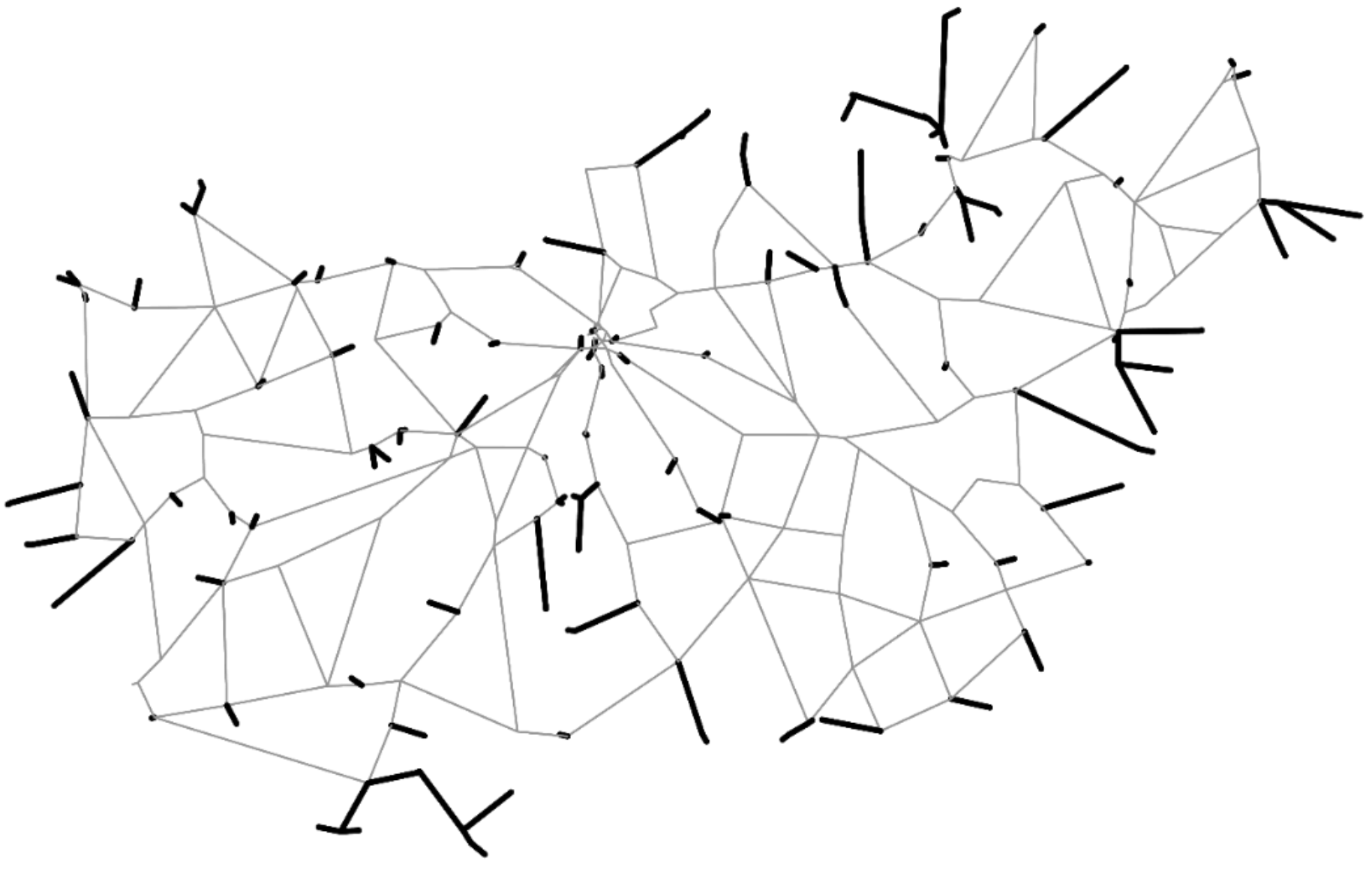

Figure 2. Line sections the disruption of which make stations unreachable for others.

### 4.1 *The Network Robustness Index*

The Network Robustness Index (NRI) was introduced by (Scott et al. 2006) as a global measure to quantitatively describe the overall resilience of a network against disruptions. The NRI can be calculated for all edges of the graph based on which the importance of the individual line sections can be determined.

To calculate the NRI for line section *e*, the shortest paths between all pairs of stations in the undisrupted graph have to be determined. Then, the lengths or durations of these paths have to be summed, which value is denoted by *c*.

Then, the edges representing line section *e* are deleted from the graph. Again, the shortest paths between all pairs of stations are determined and their lengths or durations are summed. This value is denoted by $c^e$. The NRI is calculated as the difference of these two values and is denoted by $q^e$:

$$q^e = c^e - c. \quad (1)$$

The difference is made in this order for $q^e$ to be non-negative since for most kinds of weights the deletion of a line section increases the sum of the weights of the shortest paths (or at least does not decreases it, but for a famous exception that occurs in flow models, the Braess-paradox see (Braess 1968)). This can be done for all line sections or for multiple line sections. If line sections *e* and *f* are simultaneously deleted, the NRI is calculated as

$$q^{ef} = c^{ef} - c. \quad (2)$$

The value of $q_{ab}^e$ (the difference in the shortest path between stations *a* and *b* in the disrupted and in the undisrupted network) shows if the shortest path in the undisrupted network passes through line section *e*. If $q_{ab}^e = 0$, then line section *e* is not part of the shortest path between stations *a* and *b* neither in the network represented by $G^0$ nor in the one represented by $G^e$. If $q_{ab}^e > 0$, then by deleting line section *e* the length or duration of the shortest path between station *a* and *b* increases compared to the shortest path in undisrupted network. This means that line section *e* was part of the shortest path in the undisrupted network but there is still a non-infinite route between stations *a* and *b* in the disrupted network.

### 4.2 *The redundancy index*

The Network Robustness Index measures the increase in the total network trip length or the total network travel time in the case of the deletion of a line section. But on the disruption of line section *f* the exact route of the shortest path between stations *a* and *b* changes compared to the shortest path in the undisrupted network.

Let us assume that the shortest path between stations *a* and *b* in $G^0$ did not pass through line section *e* but in $G^f$ it does. How much would be the additional increment in the shortest path if *e* would be deleted, too? In other words, how much would the shortest path between stations *a* and *b* be longer in the $G^{ef}$ graph than it is in the $G^f$ graph, i.e., we want to know how much total increase is caused by deleting not only line section *f* but also line section *e* for those paths that did not pass through line section *e* in graph $G^0$ but did pass through in in graph $G^f$. This increase is the redundancy provided by line section *e* to line section *f*. Paths that pass through line section *e* neither in graph $G^0$ nor in graph $G^f$ or pass through it in both are not relevant, since they are not sensitive for the disruption of line section *e*.

Therefore, only those shortest paths are taken into account for which $q_{ab}^e = 0$. The $r^{ef}$ redundancy index is defined by the sum of the increase of the shortest paths in $G^{ef}$ compared to the sum of the shortest paths in $G^f$:

$$r^{ef} = q^{ef} - q^f = \left(c^{ef} - c\right) - \left(c^f - c\right) = c^{ef} - c^f. \quad (3)$$

By calculating $r^{ef}$ for all *f* line sections that are not identical with *e* and summing them up one gets the total redundancy that line section *e* provides to line section *f*:

$$r^e = \sum_f r^{ef} = \sum_f \left(q^{ef} - q^f\right) = \sum_f \left(c^{ef} - c^f\right). \quad (4)$$

This definition was introduced by (Jenelius 2010).

### 4.3 *Application on 1-edge-connected graphs*

It can be seen from the definition, that if such line section(s) are deleted from the graph that make at least one station unreachable from the others, the value of both $q^e$ and $r^e$ becomes infinity. The railway network of Hungary has this property, which means that the graph describing it is a so-called 1-edge-connected graph. In several cases, by deleting only one line section from the $G^0$ graph the graph will remain connected.

However, if two line sections are deleted, the number of reasonable results will rapidly decrease. If all these line sections were excluded from the calculations, only a few would remain and if only those line sections were excluded which give infinity as a result in that particular calculation, then different line section would be taken into account for each $f$ line section, which would make the obtained $r^e$ values incomparable to each other.

Therefore, it is practical to use the reciprocals of the travel time and trip length values of the shortest paths. By changing the order in which the difference is calculated in the summation of Equation 2, the redundancy index remains positive since longer distances mean shorter values in the reciprocal space.

By summing the values of the redundancy indices calculated in the reciprocal space for all $f$ line sections one gets the total redundancy of a line section $e$:

$$\sum_f r_\ell^{ef\prime} = \sum_f \left(c_\ell^{f\prime} - c_\ell^{ef\prime}\right) = \sum_f \left(\sum_{\langle a,b\rangle} \frac{1}{\ell_{ab}^{f}} - \sum_{\langle a,b\rangle} \frac{1}{\ell_{ab}^{ef}}\right), \tag{5}$$

$$\sum_f r_t^{ef\prime} = \sum_f \left(c_t^{f\prime} - c_t^{ef\prime}\right) = \sum_f \left(\sum_{\langle a,b\rangle} \frac{1}{t_{ab}^{f}} - \sum_{\langle a,b\rangle} \frac{1}{t_{ab}^{ef}}\right). \tag{6}$$

However, it is more informative to normalize these values with values of the total trip length or the total travel time of the undisrupted network (which value is denoted by $c_\ell'$ and $c_t'$, respectively):

$$r_\ell^{e\prime} = \frac{\sum_f r_\ell^{ef\prime}}{c_\ell'} = \frac{\sum_f \left(c_\ell^{f\prime} - c_\ell^{ef\prime}\right)}{c_\ell'} = \frac{\sum_f \left(\sum_{\langle a,b\rangle} \frac{1}{\ell_{ab}^{f}} - \sum_{\langle a,b\rangle} \frac{1}{\ell_{ab}^{ef}}\right)}{\sum_{\langle a,b\rangle} \frac{1}{\ell_{ab}^{0}}}, \tag{7}$$

$$r_t^{e\prime} = \frac{\sum_f r_t^{ef\prime}}{c_t'} = \frac{\sum_f \left(c_t^{f\prime} - c_t^{ef\prime}\right)}{c_t'} = \frac{\sum_f \left(\sum_{\langle a,b\rangle} \frac{1}{t_{ab}^{f}} - \sum_{\langle a,b\rangle} \frac{1}{t_{ab}^{ef}}\right)}{\sum_{\langle a,b\rangle} \frac{1}{t_{ab}^{0}}}. \tag{8}$$

The $r^{e\prime}$ redundancy index is the total relative decrease in the reciprocal trip length or travel time for those shortest paths that do not pass through the line section $e$ in the undisrupted network but pass through it in the case of the disruption of line section $f$ with line section $e$ fixed for the calculation.

However, because of the definition, the redundancy values of line sections calculated in a specific graph cannot be used to compare with values obtained for line sections in other graphs. They have meaning only in that specific graph that they were calculated for and also only relative to the redundancy values of other line sections.

## 5 LINE SECTIONS PROVIDING REDUNDANCY FOR THE RAILWAY NETWORK OF HUNGARY

Calculating the redundancy values for all the line sections in the railway network of Hungary the network elements that that provide an alternative route with the smallest increase in the length or duration of the shortest paths are identifiable. The numbering of railway lines according to the Hungarian State Railways (Magyar Államvasutak, MÁV) is used to refer to the lines, which accessible also at (VPE2019).

### 5.1 *Minimal trip lengths*

By calculating the value of the $r_\ell^{e\prime}$ index for all $e$ line sections in graph $G_\ell$, the results mapped in Figure 3 are obtained.

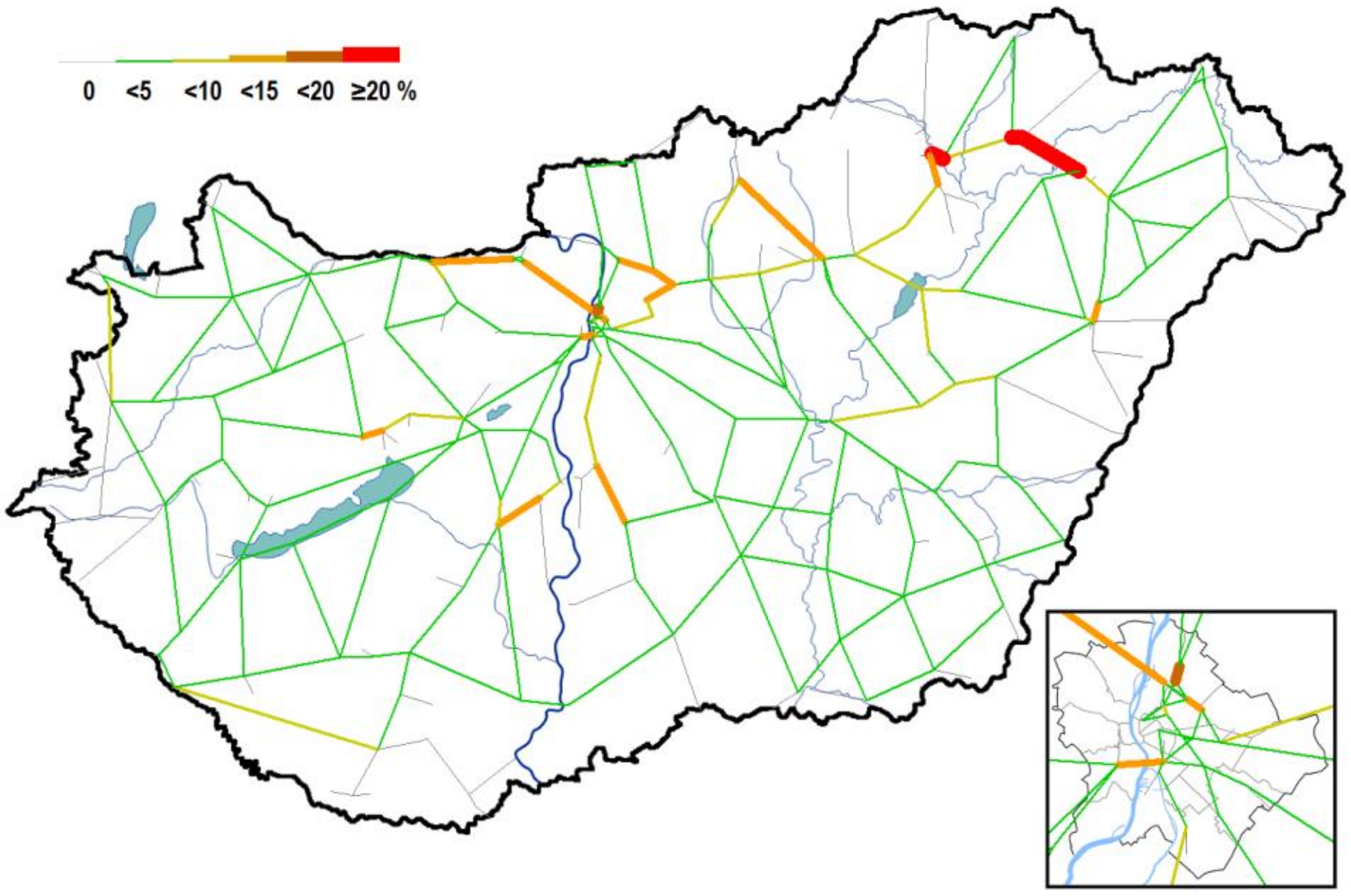

Figure 3. Redundancy values of the line section of the railway network of Hungary for paths with minimal trip length.

The results show that the electrified single tracked section between stations Görögszállás and Mezőzombor on line 80 has the highest redundancy value with 32%. The alternative routes using it have the highest relative increase in their total length were this line section unusable. The Tisza bridge at Tokaj is part of this line section which, though not thought as a particularly important crossing (compared to the Tisza bridge at Szolnok), is, in fact, vital for the network due to its high redundancy value.

The line sections with the next two highest redundancy value are also part of line 80: the Szerencs–Mezőzombor ($r_\ell'$ = 27%) and the Felsőzsolca–Miskolc-Tiszai ($r_\ell'$ = 22%) line sections. These three line sections are part of the same Miskolc–Nyíregyháza line section, which connects the Budapest–Debrecen–Nyíregyháza line (line 100/100a) and the Budapest–Hatvan–Miskolc line (line 80a/80), both part of the Trans-European Transport Networks and handling heavy freight traffic. In the case of the disruption of these main lines, the Miskolc–Nyíregyháza line section in question is the shortest, and therefore, the most practical direction to reroute the traffic.

Lines 2 and 4 between stations Angyalföld and Almásfüzitő also have a significant 14% redundancy. This redundancy value is (or, in fact, were to) given mainly to the Southern Railway Bridge at Budapest. Though in theory paths through these lines are good alternatives being only slightly longer than the section of line 1 between Budapest and Almásfüzitő (trough Tatabánya), in reality, it is not true due to not only the poor condition of the tracks of line 4 between Dorog and Almásfüzitő but also because being single tracked and not electrified in its entire length.

On the contrary, the 13% redundancy value of the single tracked but electrified line 77 between stations Galgamácsa and Vácrátót is in good agreement of the original purpose of this line: to provide a short rerouting alternative for the freight transport directed towards Western Europe through Slovakia in the case of the disruption of the main lines in and around Budapest. However, to fully utilize the advances of this route, the development of the connecting Vác–Vácrátót section of line 71 and the Galgamácsa–Aszód section of line 78 is also necessary.

At this point, the dangers in the decrease in the number of diesel locomotives has to be pointed out. As the electrification of the railway lines goes on, less and less diesel engines are needed. However, the electric network is also a critical infrastructure. Either in the case of the disruption of the overhead lines or disturbances in the electricity supply the results are the same as if the tracks themselves were damaged if there is no sufficient number of diesel locomotives to take over. Because of this, even though using electric locomotives under normal operational conditions is much cheaper than using diesels, the lack of enough usable diesel engines can have severe economic consequences.

## 5.2 *Minimal travel times*

The values of index $r_t^{e\prime}$ for all $e$ line sections in graph $G_t$, are mapped in Figure 4.

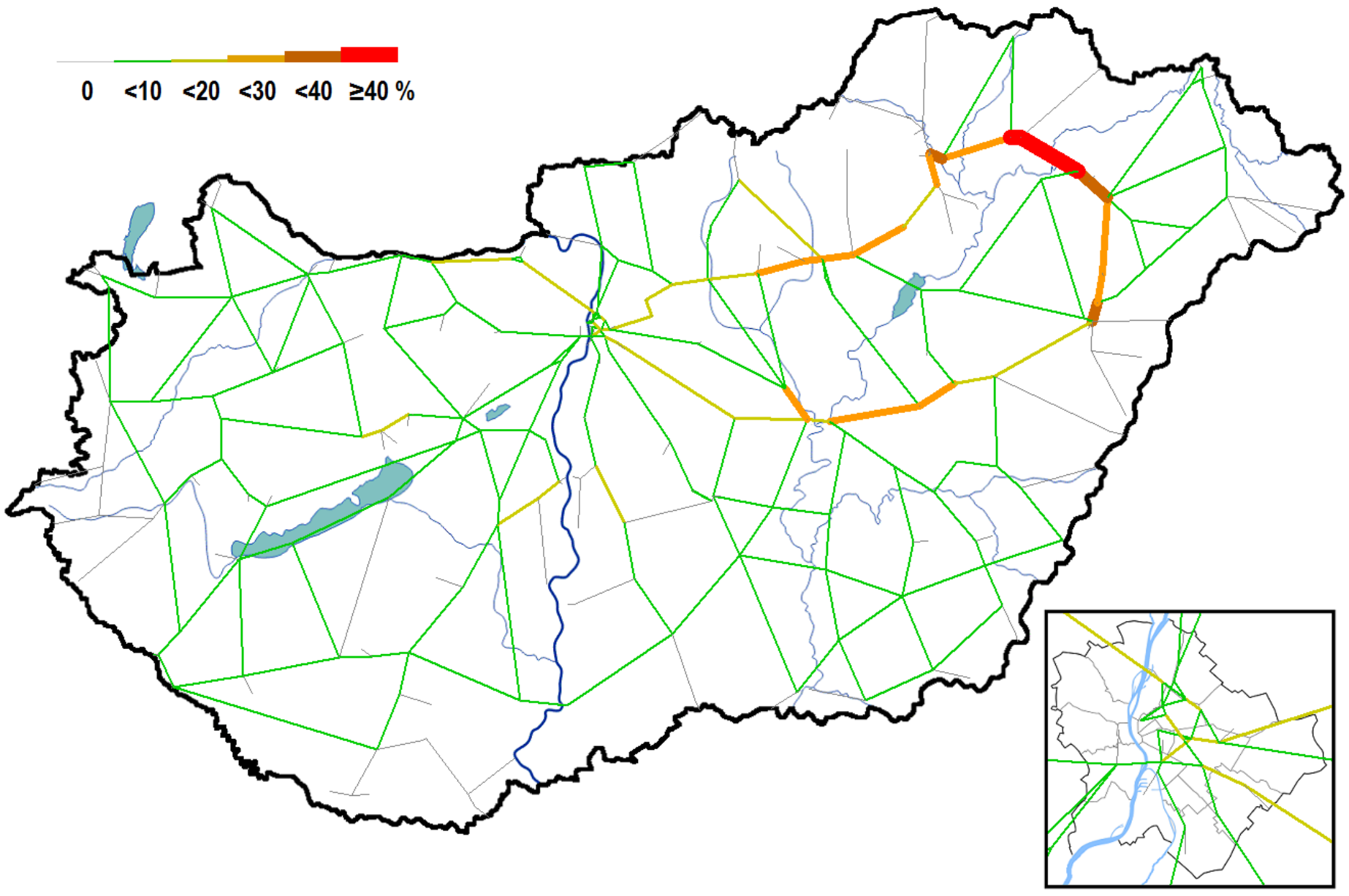

Figure 4. Redundancy values of the line section of the railway network of Hungary for paths with minimal travel time.

The results are similar for line 80: the line section that has the highest redundancy ($r_t' = 48\%$) is the line section between Görögszállás and Mezőzombor. The neighboring line sections also show high redundancy: the value of $r_t'$ for the Mezőzombor–Szerencs line section is 42%, for the Felsőzsolca–Miskolc-Tiszai line section 36%, and for the Görögszállás–Nyírtelek line section 31%. Therefore, the part of line 80 between Miskolc and Nyíregyháza is the most effective alternative route between lines 80 and 100 both for trip lengths and travel times.

The line section between Debrecen and Apafa has 32% redundancy value, which indicates its bottleneck nature: if disrupted, the alternative routes run via the transversal lines between lines 80 and 100, which have lower line speeds and are therefore long detours.

The redundancy of lines 2 and 4 for travel times is much less pronounced as it was for trip lengths because of the conditions of line 4, but it is still not negligible. By developing line 77 between Galgamácsa and Vácrátót (and the connecting Vác–Vácrátót section of line 71 and the Galgamácsa–Aszód section of line 78), both ways could be used as an alternative for the Southern Railway Bridge.

The small redundancy value ($r_t' = 3\%$) of the Southern Railway Bridge means that the rerouting of those alternative routes in the case of the disruption of the bridge that do not pass through it in the undisrupted network do not have a large effect on the network as a whole. This is because the traffic of the Northern Railway Bridge in the undisrupted network is almost exclusively the traffic of line 2, which is so small that the rerouting of these paths towards Baja (in spite of the very large increase in the travel time) does not affect much the performance of the network as a whole. For the paths crossing the Danube at Baja in the undisrupted network, the detour via Budapest increases the travel time so much that crossing the Danube at either through the Southern or Northern Railway Bridge does not make a big difference.

# 6 CONCLUSION

The railway network of Hungary is one of the densest in the world. The benefits, however, can only be reaped if the infrastructure is maintained and developed properly. This has to include not only the lines with heavy traffic but also the lines with high redundancy, even if these lines have little traffic under normal operational conditions, in order to provide sufficient alternative routes in the case of disruptions.

The lines with high redundancy include the section of line 80 between Miskolc and Nyíregyháza, which is not only part of the RFC corridor 6 (Mediterranean) but also the connection between its two branches in Hungary. Being single-tracked in most of its length, this is the line section that has to be developed first by building a second track to be able to handle the increased rerouted traffic were lines 80 or 100 disrupted.

Paths through lines 2 and 4 are potential alternatives for the Southern Railway Bridge in Budapest, through which all of the international freight transport crosses the Danube in Hungary being the common network element of RFC corridors 6 (Mediterranean) and 7 (Orient). By building a second track and electrifying them in their whole length can provide a possible alternative route for the only double-tracked and the only electrified bridge across the Danube in the country.

By building a second track for line 77 (and its connecting lines), these lines can also be an alternative of the Southern Railway Bridge by rerouting the traffic directed to Western Europe.